\documentclass[twocolumn,showpacs,preprintnumbers,amsmath,amssymb]{revtex4}
\usepackage{graphicx}% Include figure files
\usepackage{dcolumn}% Align table columns on decimal point
\usepackage{bm}% bold math

\begin{document}
\title{Test of the Wiedemann-Franz law in an optimally-doped cuprate}
\author{Romain Bel$^{1}$, Kamran Behnia$^{1}$, Cyril Proust$^{2}$, Peter van der Linden$^{3}$,
 Duncan Maude$^{3}$, Sergey I. Vedeneev$^{3,4}$}
 \affiliation{(1)Laboratoire de Physique Quantique(CNRS), ESPCI, 10 Rue de Vauquelin,
 75231 Paris, France \\
 (2)Laboratoire National des Champs Magn\'etiques Puls\'es(CNRS), BP
4245, 31432 Toulouse, France \\
(3)Grenoble High Magnetic Field Laboratory(CNRS-MPI), BP 166, 38042 Grenoble, France \\
(4)Lebedev Physical Institute, Russian Academy of Sciences, 117924
Moscow, Russia}

\date{February 17, 2004}

\begin{abstract}
We present a study of heat and charge transport in
Bi$_{2+x}$Sr$_{2-x}$CuO$_{6+\delta}$ focused on the size of the
low-temperature linear term of the thermal conductivity at optimal
doping level. In the superconducting state, the magnitude of this
term implies a d-wave gap with an amplitude close to what has been
reported. In the normal state, recovered by the application of a
magnetic field, measurement of this term and residual resistivity
yields a Lorenz number $L = \kappa_{N}\rho_{0}/T = 1.3 \pm 0.2
L_{0}$. The departure from the value expected by the
Wiedemann-Franz law is thus slightly larger than our estimated
experimental resolution.
\end{abstract}


\maketitle

In spite of many years of intense research by a sizeable fraction
of the condensed-matter physics community, high T$_c$
superconductivity remains a mystery. A central question is the
extent of the validity of Landau's Fermi liquid picture to
describe the elementary excitations of the ground state.
High-T$_c$ cuprates are doped Mott insulators and are host to a
particularly strong Coulomb repulsion neglected in the Fermi
liquid picture. It remains to be established, however, to what
extent this becomes an obstacle for the formation of Landau
quasi-particles in the T=0 limit. A recent attempt to answer this
question has been made by measuring the subkelvin thermal
conductivity of the normal state in order to check for the
validity of the Wiedemann-Franz law which is a robust signature of
a Fermi liquid\cite{hill,proust,nakamae}. The validity of this
universal law relating the magnitude of thermal and electrical
conductivities is expected in the T=0 limit, disregarding the fine
details of electronic scattering and Fermi surface. On the other
hand, various scenarios based on electron fractionalization lead
to its violation.

In this paper, we present a first experimental study to check the
validity of the Wiedemann-Franz(WF) law in a hole-doped cuprate at
optimal-doping concentration. Recovering the normal state of
Bi$_{2+x}$Sr$_{2-x}$CuO$_{6+\delta}$ (Bi-2201) by the application
of a strong magnetic field, we found a thermal conductivity
slightly larger than what is expected by the WF law. The departure
appears to be genuine as it is significantly larger than the
experimental precision obtained in the verification of the WF law
in a simple metal.

In-plane thermal conductivity ($\kappa$) in presence of a magnetic
field applied along the c-axis was measured with a standard
two-thermometers-one-heater set-up which permitted to measure the
in-plane electric resistivity in the same conditions. Bi-2201
single crystals, with typical dimensions of (2-8) $\times$
(400-800) $\times$ (600-900)$\mu m ^{3}$, were grown in a gaseous
phase in closed cavities of a KCl solution melt as detailed
elsewhere\cite{vedeneev}.  Doping level in stoichiometric
Bi$_{2}$Sr$_{2}$CuO$_{6+\delta}$ can be modified by replacing
Sr$^{2+}$ either with Bi$^{3+}$ or with
La$^{3+}$\cite{maeda,harris}. We succeeded to make single-phase
high quality crystals of Bi$_{2+x}$Sr$_{2-x}$CuO$_{6+\delta}$ in
the range of 0.17$<$x$<$0.7. The maximum [resistive] $T_{c}$ found
in this system was $\sim$ 10.5 K in agreement with previous
studies\cite{vedeneev,harris}. In
Bi$_2$Sr$_{2-x}$La$_{x}$CuO$_{6+\delta}$, on the other hand, the
maximum $T_c$ is 38K \cite{ono}. Thus, the disorder associated
with (Sr,Bi) substitution is apparently stronger than the one due
to (Sr,La) doping and leads to a lower $T_c$ and a higher residual
resistivity in Bi$_{2+x}$Sr$_{2-x}$CuO$_{6+\delta}$ compared to
Bi$_2$Sr$_{2-x}$La$_{x}$CuO$_{6+\delta}$\cite{ono2}.

% Figure Hall data
\begin{figure}
\resizebox{!}{0.35\textwidth}{\includegraphics{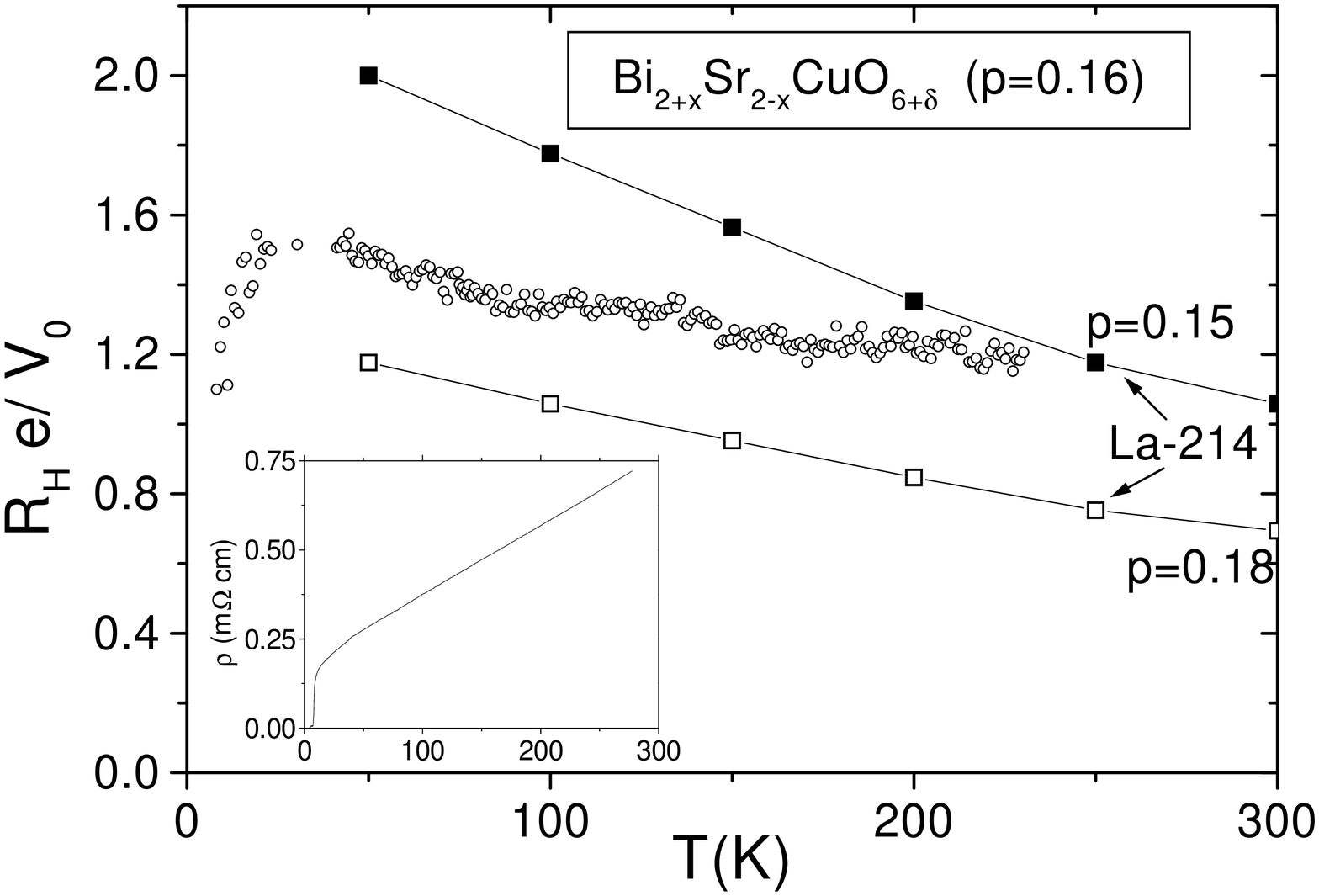}}
\caption{\label{fig1} Renormalised Hall coefficient in a
Bi$_{2+x}$Sr$_{2-x}$CuO$_{6+\delta}$ sample compared with the
values reported for La-214\cite{hwang} for two different doping
levels. Inset shows the temperature-dependence of resistivity.}
\end{figure}

An unambiguous determination of carrier concentration,\emph{p}, in
high-T$_{c}$ cuprates other than
La$_{2-x}$Sr$_{x}$Cu$_{2}$O$_{4}$(La-214) is not straightforward.
However, by comparing Hall coefficients of various families of
cuprates, Ando \emph{et al.}\cite{ando} have shown that the
magnitude of renormalised Hall coefficient($R_{H}e/V_{0}$, where
$V_{0}$ is the volume associated with each Cu atom) is an
appropriate measure of carrier concentration. In order to check
the doping level of our samples we measured the Hall coefficient
in a number of them and found that the magnitude of $R_{H}e/V_{0}$
is close to what is expected for an optimal doping cuprate. As
seen in Fig. 1 which presents a typical curve, the magnitude of
renormalised Hall coefficient is somewhere between the values
obtained for La-214 at \emph{p}=0.15 and \emph{p}=0.18. A similar
result has been recently reported by Ono and Ando\cite{ono2}.
Together with the linear resistivity observed from room
temperature down to $T_{c}$ (see the inset of Fig.1), this
provides compelling evidence that the physics of cuprates at
optimal doping level can be explored in Bi-2201, which contrary to
other hole-doped cuprates, presents a resistive upper critical
field within the range of available DC magnetic
fields\cite{vedeneev}.

As the required field ($\sim$ 25 T), however, still exceeds the
range provided by available commercial magnets, the experiment was
performed in a resistive magnet at Grenoble High Magnetic Field
Laboratory. Measuring subkelvin thermal conductivity in this
context faced two major technical challenges. The first was to
cool down the sample and thermometers held in vacuum to low
temperatures in spite of strong mechanical vibrations associated
with the circulation of cooling water. This problem was partially
solved by designing a new insert with a vacuum chamber containing
the entire thermal conductivity set-up which was placed inside the
mixing chamber of a Kelvinox top-loading dilution fridge\cite{van
der linden}. The second was to measure accurately the temperature
at high magnetic fields given the non-negligible
magneto-resistance of the RuO$_{2}$ thermometers used in the
set-up and the absence of any zero-field zone. This second problem
was resolved by using Coulomb Blockade
Thermometry(CBT)\cite{pekola}. An array of 100 tunnel junctions
provided by Nanoway (Finland) was attached to the cold finger and
used as a field-independent thermometer\cite{kauppinen} for
calibrating the shift in the resistance of RuO$_{2}$ thermometers
with the application of magnetic field.

% Figure kappa at H=0
\begin{figure}
\resizebox{!}{0.7\textwidth}{\includegraphics{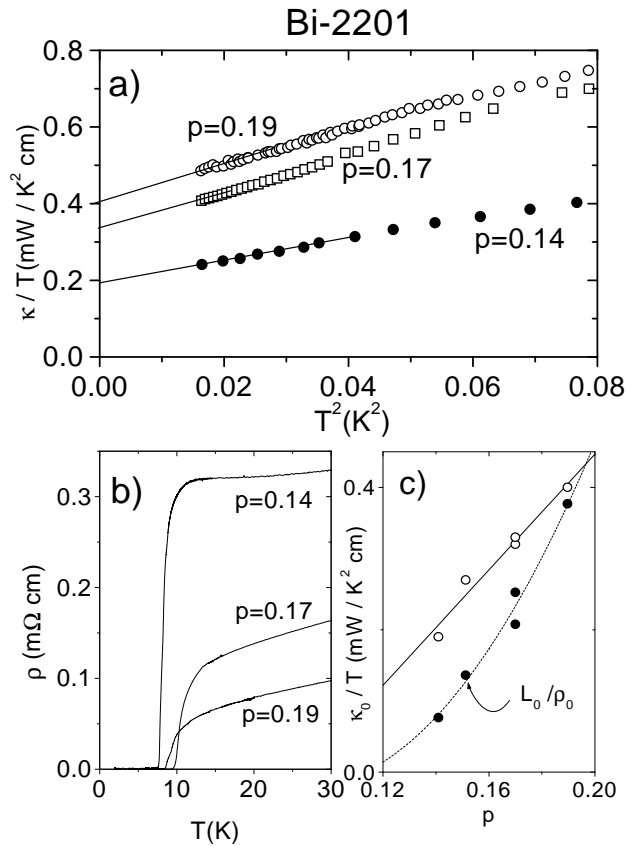}}
\caption{\label{fig2} a) Subkelvin thermal conductivity in three
different Bi-2201 single crystals. b) Temperature dependence of
the zero-field resistivity in the three samples. c) A comparison
between the doping dependence of $\kappa_0/T$ (open circles) and
of L$_{0}$/$\rho_{0}$ (solid circles). Lines are guides for eye.}
\end{figure}

\emph{Zero magnetic field:} Fig. 2(a) displays the temperature
dependence of thermal conductivity for three crystals of Bi-2201
at zero magnetic field. Since the lattice thermal conductivity is
expected to display a cubic temperature dependence at very low
temperatures, ($\kappa/T$) is plotted as a function of $T^2$ in
order to extract a linear term, $\kappa_0/T$, associated with the
electronic contribution. The magnitude of this linear term varies
from 0.19 in the most underdoped sample (\emph{p}=0.14) to 0.40
mW/K$^2$cm in the most overdoped sample (\emph{p}=0.19) (See fig.
2(c)). Doping level in these samples was estimated using the
reported empirical relationship between $T_{c}$ and carrier
concentration in Bi-2201\cite{ando}. Fig. 2(b) presents the
temperature dependence of resistivity in the same samples.

\begin{table}
  \centering
\begin{tabular}{|c|c|c|c|c|c|}

\hline Compound& T$_c$(K) &$\kappa_0/T$ & $\frac{v_F}{v_2}$
&$\Delta_0$(meV)&$\Delta_{tunnel}$(meV)\\ \hline

Bi-2212 & 89 & 0.15 &19 & 30 & $\sim$40\cite{renner}\\
\hline

Bi-2201 & 11 & 0.33 &35 & 16 & 12-15\cite{kugler,vedeneev2}\\
\hline
\end{tabular}
  \caption{A comparison of two Bi-based cuprates at optimal doping level.
$\kappa/T$ is expressed in mW/K$^2$cm units. $\Delta_0$ was
calculated using v$_F$=250 km/s and k$_F$=0.7$\AA^{-1}$ obtained
in ARPES studies on Bi-2212\cite{mesot}.}\label{T1}
\end{table}

A finite $\kappa_0/T$ is a consequence of heat transport by nodal
quasi-particles of the \textit{d}-wave superconducting gap. It has
been already reported in a variety of hole-doped
cuprates\cite{taillefer,nakamae2,chiao,takeya,sutherland}.
According to the theory of transport in a \textit{d}-wave BCS
superconductor\cite{graf,durst}, the magnitude of $\kappa_0/T$ is
intimately related to the fine structure of the superconducting
gap. Quantitatively, the theory states that $\kappa_{0}/T
=\frac{{k_{B}}^{2}}{3\hbar}\frac{{v_{F}}}{{v_{2}}}\frac{n}{d}$\cite{durst}.
Here, $v_F$ and $v_2$ are velocities of nodal quasi-particles
normal and parallel to the Fermi surface and $\frac{n}{d}$ is the
stack density of CuO$_2$ planes. According to the theory, the
magnitude of $\kappa_0/T$ is universal in the sense that it does
not depend on the scattering rate\cite{graf}. This prediction is
valid in the clean limit, that is, when $\gamma \ll k_{B}T_c$,
with $\gamma$ being the impurity bandwidth. Experimentally, both
in YBa$_{2}$Cu$_{3}$O$_{6.9}$(Y-123)\cite{taillefer} and in
Bi$_{2}$Sr$_{2}$CaCu$_{2}$O$_{8+\delta}$(Bi-2212)\cite{nakamae2},
the magnitude of $\kappa_0/T$ has been found to be independent of
impurity concentration up to the highest scattering rates
investigated for optimally-doped samples. On the other hand,
according to recent studies on La-214\cite{takeya,sutherland} and
on Y-123\cite{sutherland}, $\kappa_0/T$ is a monotonously
increasing function of doping
concentration\cite{takeya,sutherland}.

Fig.2(c) confirms the latter behavior in the case of Bi-2201. In
the limited range of our exploration, the magnitude of
$\kappa_0/T$ increases with the increase in doping level. This
allows us to determine the magnitude of $\kappa_0/T$ in Bi-2201 at
optimal doping level(\emph{p}=0.17) to be 0.33 mW/K$^2$cm. This
value is more than twice the magnitude reported for
Bi-2212\cite{nakamae2,chiao}, which has a substantially higher
$T_c$. This is not surprising, since $\kappa_0/T$ is a zero-energy
probe inversely proportional to the superconducting
gap\cite{sutherland}. Now, it is tempting to forget that Bi-2201
is \emph{not} in the clean limit (as defined above) and compute
$v_F/v_2$=35, using the procedure already used for other
cuprates\cite{chiao,nakamae2,sutherland}. Since $v_2$ is
proportional to the angular slope of the gap at a nodal position,
this number can be related to the magnitude of the superconducting
gap assuming a regular \textit{d}-wave angular
dependence($\Delta=\Delta_0\cos$2$\phi$). This yields $\Delta_0 =
\hbar k_F v_2 / 2 $ = 16 meV comparable with the magnitude
reported by tunnelling studies\cite{kugler,vedeneev2} (See table
I). Note that while $T_c$ is suppressed by a factor of 8 in
Bi-2201, the associated reduction in the magnitude of the
superconducting gap is considerably smaller and the magnitude of
$\kappa_0/T$ appears to reflect this latter trend. This leads to
an exceptionally high $\Delta$/k$_{B}$T$_{c}$ ratio in Bi-2201
($\sim$15)\cite{kugler}, 7 times larger than the value expected
for a \textit{d}-wave BCS superconductor.

\emph{Normal state:} Now, we turn our attention  to the issue of
the validity or violation of the Wiedemann-Franz law, which states
that the Lorenz number, $L = \kappa_{N}\rho_{0}/T$, should be
equal to Sommerfeld's value:

\begin{equation}\label{E1}
L_0=\frac{\pi^{2}}{3}\frac{k_{B}^{2}}{e^{2}}=24.4\times10^{-9}
V^{2}/K^{2}
\end{equation}

Here $\kappa_{N}/T$ and $\rho_{0}$ are thermal conductivity and
electric resistivity of the normal state in the \textit{T=0}
limit. Before presenting the data on heat conductivity in the
normal state, let us note that in our samples the residual thermal
conductivity in the \emph{superconducting} state exceeds the
magnitude of the thermal conductivity in the \emph{normal }state
expected by the magnitude of the resistivity (either $\rho (T_c)$
or $\rho_{0}$) and according to the WF law. In other words,
$\kappa_0/T
>$ $L_{0}/\rho_{0}$. This surprising inequality was first reported
in the case of underdoped La-214\cite{takeya}. In Bi-2201, as seen
in Fig. 2(c), it persists at optimal doping, but gradually
disappears with increasing p. In this context, a field-induced
enhancement of $\kappa_0/T$, like the one observed in
optimally-doped Bi-2212\cite{aubin}, would lead to a strong
violation of the WF law. In the underdoped La-214, on the other
hand $\kappa_0/T$ decreases with the application of a magnetic
field\cite{sun,hawthorn}.

As seen in Fig. 3, however, the application of a magnetic field,
strong enough to destroy any trace of superconductivity in charge
transport, has little effect on low-temperature thermal
conductivity of Bi-2201. In the slightly overdoped sample
(\emph{p}=0.19), the magnetic field leads to a slight increase in
the residual linear term. No change in $\kappa/T$ is observed for
H=15 T and H=20 T. The few available data points for H=25 T
confirms that a reliable linear term for this field can be
extracted from the data obtained at lower fields. Assuming that
the magnetic field does not affect the slope of $\kappa/T$ in the
ballistic regime, we estimate $\kappa_{N}/T$ =0.46$\pm$0.04
mW/K$^{2}$cm by extrapolating the H=20 T data to \textit{T=0}
along a line parallel to the H=0 slope. In the case of the
optimally-doped sample (p=0.17), a magnetic field of 25 T leaves
the thermal conductivity virtually unchanged in the explored
temperature range. Thus, the zero-field extrapolation for this
sample ($\kappa_{0}/T$ =0.32$\pm$0.03 mW/K$^{2}$cm) may be
considered as the magnitude of $\kappa_{N}/T$. As seen in the
figure,  for both samples, $\kappa_{N}$/T exceeds
$L_{0}$/$\rho_0$. The excess is about 20(30) percent for the
\emph{p}=0.19(0.17) sample.

% Figure kappa in a field
\begin{figure}
\resizebox{!}{0.7\textwidth}{\includegraphics{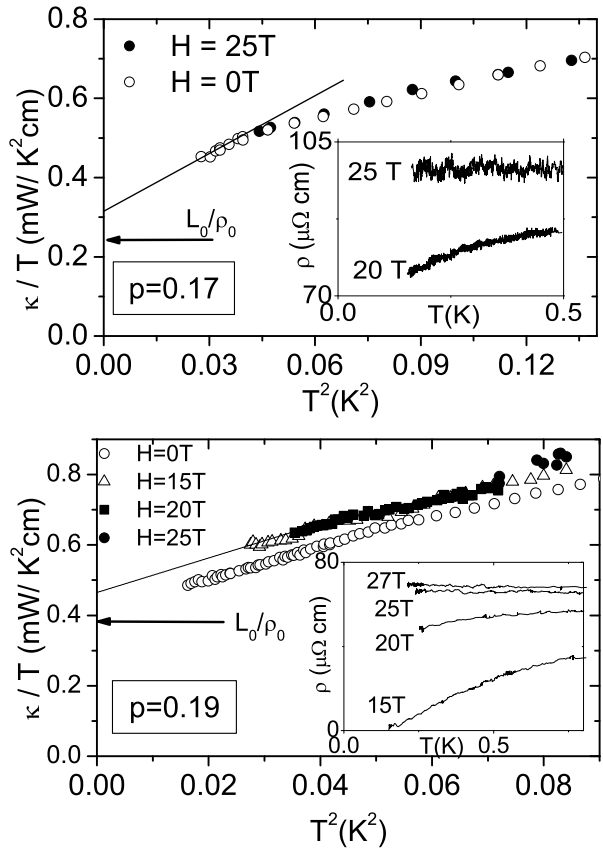}}
\caption{\label{fig3}  Upper and lower panels: The effect of
magnetic field on subkelvin thermal conductivity in two Bi-2201
samples with different doping levels. Solid lines represent the
extrapolated low-temperature behavior and arrows show the expected
WF magnitude. The insets display the resistivity data for the two
samples. }
\end{figure}

In order to estimate our resolution in the verification of the WF
law in a simple metal, we used the same set-up to measure the
low-temperature thermal and electric conductivities of a 17
$\mu$m-diameter gold wire at H=0 and H=25 T\cite{van der linden}.
The application of the magnetic field led to a three-fold increase
in $\rho_0$ and a concomitant decrease in $\kappa/T$ of the gold
wire. Using Coulomb blockade thermometry, we succeeded in
verifying the WF law with a precision of 1(3) percent at H=0 (25
T)\cite{van der linden}. Thus, the deviation observed in Bi-2201
is much larger than our experimental precision on the absolute
magnitude of $\kappa/T$.

Besides absolute thermometry, other sources of error may arise in
the case of sub-millimetric and anisotropic samples. The geometric
factor, assumed identical for $\kappa$ and $\rho$ in our analysis,
may be different for electric and thermal transport. Since we did
not observe any correlation between the magnitude of thermal
conductivity and the width of gold electrodes, we can exclude the
finite width of the contacts as a significant source of
discrepancy. An eventual \textit{c}-axis contamination of the
presumably in-plane conductivity would lead to an overestimation
of $\rho_0$. Moreover, as the electric transport is much more
anisotropic than heat transport, the measured $\kappa$/T would be
less affected by such a contamination.  We estimate that in the
case of \emph{p}=0.17 sample (with a thickness of $2.5 \mu m$),
this can lead to an error of 6 percent in the absolute magnitude
of residual resistivity.

The sum of the three identified sources of experimental error
(extrapolation to zero temperature, absolute thermometry and
overestimation of residual resistivity due to a \textit{c}-axis
contamination) yield an uncertainty of 17 percent. We can
therefore conclude that for the p=0.17 sample, the Lorenz number
exceeds Sommerfeld's value by a slight yet significant margin,
that is $L = 1.3 \pm 0.2 L_{0}$.

Prior to this work, compelling evidence for the validity of the WF
law in the overdoped regime was reported for overdoped
Tl-2201(\emph{p}=0.26)\cite{proust} and for heavily overdoped
La-214(\emph{p}=0.30)\cite{nakamae}. Together with the detection
of a purely $T^2$ temperature-dependence of the resistivity in the
latter compound, this seems to establish that the ground state in
the overdoped limit is indeed a Fermi liquid.  A previous study
performed on an electron-doped cuprate at optimal doping level,
pointed to a different conclusion\cite{hill}. Note, however, that
in the latter experiment the extraction of $\kappa_{N}/T$ was
complicated by the presence of an intriguing downturn in thermal
conductivity below 0.3 K. For temperatures above 0.3 K, Hill
\emph{et al.}, extracted a $\kappa_{N}/T$ which exceeded the
expected WF value by a factor of 1.7\cite{hill}. Our results do
not suffer from the presence of this downturn which was also
observed in overdoped La-214\cite{nakamae,sutherland}. They yield
a $\kappa_{N}/T$, still larger than, but closer to, the WF value.
Very recently, we have  found that the slight deviation observed
here is considerably enhanced with underdoping and/or
disorder\cite{proust2}. This indicates that the slight departure
from the WF law at optimal doping level is not due to an
experimental imperfection.

The outcome of this study has interesting implications for the
debate on the origin of the anomalous transport properties of
cuprates. In theories invoking electron fractionalization in
cuprates\cite{senthil}, the zero-temperature excitations of the
normal state are not Landau quasi-particles and the WF law is not
expected to be valid. In the case of a Luttinger liquid, for
example, a strong violation of the WF law has been theoretically
predicted\cite{kane}. On the other hand, if the anomalous
properties of the normal state is due to the existence of a
competing hidden order, then the validity of the WF law at T=0 is
still expected\cite{kim}.

In summary, we have measured heat transport in the normal and
superconducting state of Bi-2201. In the normal state, we have
found a Lorenz number slightly but significantly larger than
Sommerfeld's value.

We thank L. Taillefer for fruitful discussions and C. Gianese for
technical assistance.

\end{document}